\documentclass{Interspeech}

\interspeechcameraready

\title{Lightweight and Robust Multi-Channel End-to-End Speech Recognition with Spherical Harmonic Transform\thanks{$^{*}$Corresponding authors. This work is partly supported by Guangxi Science and Technology Project (2022AC16002) and National Natural Science Foundation of China
(62466055). The code is released at https://github.com/thu-spmi/CAT/blob/master/docs/whatsnew.md}}

\author[affiliation={1}]{Xiangzhu}{Kong}
\author[affiliation={1}]{Hao}{Huang$^{*}$}
\author[affiliation={2}]{Zhijian}{Ou$^{*}$}

\affiliation{School of Computer Science and Technology}{Xinjiang University}{China}
\affiliation{Speech Processing and Machine Intelligence (SPMI) Lab}{Tsinghua University}{China}

\email{
huanghao@xju.edu.cn, ozj@tsinghua.edu.cn
}

\usepackage{comment}
\usepackage{booktabs}
\usepackage{adjustbox}
\usepackage{array}
\usepackage{makecell}
\usepackage{multirow}

\begin{document}

\maketitle

\begin{abstract}
    
This paper presents SHTNet, a lightweight spherical harmonic transform (SHT) based framework, which is designed to address cross-array generalization challenges in multi-channel automatic speech recognition (ASR) through three key innovations. First, SHT based spatial sound field decomposition converts microphone signals into geometry-invariant spherical harmonic coefficients, isolating signal processing from array geometry. Second, the Spatio-Spectral Attention Fusion Network (SSAFN) combines coordinate-aware spatial modeling, refined self-attention channel combinator, and spectral noise suppression without conventional beamforming. Third, Rand-SHT training enhances robustness through random channel selection and array geometry reconstruction. The system achieves 39.26\% average CER across heterogeneous arrays (e.g., circular, square, and binaural) on datasets including Aishell-4, Alimeeting, and XMOS, with 97.1\% fewer computations than conventional neural beamformers.
\end{abstract}

\vspace{-0.25cm}
\section{Introduction}
\vspace{-0.15cm}
Recent advancements in multi-channel end-to-end automatic speech recognition (ASR) have demonstrated superior performance over single-channel systems by effectively leveraging spatial cues captured by microphone arrays \cite{chang2019mimo,chang21_interspeech, zhang2022end, yu2023mfcca,sharma2022spatial}.

Three principal approaches have emerged for spatial information utilization: conventional spatial filtering \cite{zhang2022end,kong2024cusidearray,an2022exploiting}, neural spatial modeling \cite{chang21_interspeech, yu2023mfcca, wang2022css, yoshioka2018multi}, and spherical harmonic field encoding \cite{pan2024efficient,pan2024innovative}.
Traditional beamforming methods optimize spatial filters using signal priors like delay-and-sum \cite{Klemm2008das} and minimum variance distortionless response (MVDR) \cite{Souden2009mvdr} beamformers.
For instance, EaBNet \cite{li2022eabnet} directly estimates beamforming filter coefficients via neural networks for noise suppression, while CUSIDE-Array \cite{kong2024cusidearray} employs a deep neural network (DNN) to estimate speech/noise covariance matrices that are then used to calculate MVDR filter coefficients \cite{ochiai2017multichannel}, yielding enhanced speech for streaming recognition. In contrast, neural spatial modeling sidesteps traditional signal processing constraints with architectures like TF-GridNet \cite{wang2023tf}, which integrates full-band spectral correlations and sub-band temporal dynamics via bidirectional LSTM networks, and MFCCA \cite{yu2023mfcca}, which utilizes cross-channel attention mechanisms to jointly optimize spatial localization and feature enhancement.

Despite their merits, these approaches face significant challenges. Traditional beamforming methods are often sensitive to array geometry and struggle and are less effective in real-time processing in complex environments \cite{li2022eabnet}. Meanwhile, although neural spatial modeling overcomes some of these limitations, its reliance on deep network architectures incurs substantial computational overhead, hindering practical deployment and scalability \cite{chang21_interspeech, wang2023tf}.

The spherical harmonic transform (SHT) offers a promising alternative \cite{pan2024efficient, kumar2016near}. It takes advantage of the inherent properties of spherical harmonics to project spatial sound fields onto a universal set of basis functions. This decouples the spatial representation from the specific microphone array \cite{rafaely2015fundamentals}.

This approach mitigates sensitivity to array geometry while providing a compact representation that substantially reduces computational load compared to both traditional beamforming and deep neural network methods. Recent work by Pan et al. \cite{pan2024efficient} demonstrated the efficacy of SHT in speech enhancement. They introduced the pIGCRN framework, which combines spherical harmonic coefficients with Short-Time Fourier Transform (STFT) features using dual encoders based.

Despite these advancements, the application of SHT in end-to-end ASR systems remains underexplored, especially with regard to the challenges associated with cross-array generalization. 
To bridge this gap, we present \textbf{SHTNet} — a novel, lightweight multi-channel end-to-end ASR framework based on the spherical harmonic transform (SHT), built upon the CUSIDE-Array architecture. Our framework introduces three key innovations:

\vspace{-0.1cm}
\begin{enumerate}
    \item \textbf{Unified Spatial Representation:} Enabled by the SHT, our method converts multi-channel audio signals in the element domain from various array configurations into a common spherical harmonic domain \cite{yan2011element}. Integrating spatial information into a compact representation enables the system to process data independently of the array geometry.
    \vspace{-0.1cm}
    \item \textbf{Spatio-Spectral Attention Fusion Network (SSAFN):} A hierarchical attention architecture combines coordinate-aware spatial modeling, refined self-attention channel combinator, and spectral noise suppression without conventional beamforming. Specifically, it employs JointAttention modules with CBAM \cite{woo2018cbam} and CoorAttention \cite{hou2021coordinate} for spatial and frequency feature extraction, followed by a self-attention-based channel combinator \cite{rong2022sacc} to amplify target signals, and concludes with a multi-head self-attention (MHSA) \cite{vaswani2017transformer} post-filter for residual noise suppression.
    \vspace{-0.1cm}
    \item \textbf{Rand-SHT Training Strategy:} Random channel selection and array geometry reconstruction for SHT during training, enhancing robustness to varying array geometries and environmental conditions.
\end{enumerate}
\vspace{-0.1cm}

Our approach shows superior performance in addressing array variability and environmental uncertainty, all while maintaining computational efficiency. Extensive experiments on both in-domain (ID) and out-of-domain (OOD) datasets show that our method outperforms existing benchmarks, with lower computational costs and fewer parameters.

\section{Method}

\begin{figure}[t]
    \centering
    \includegraphics[width=0.5\textwidth]{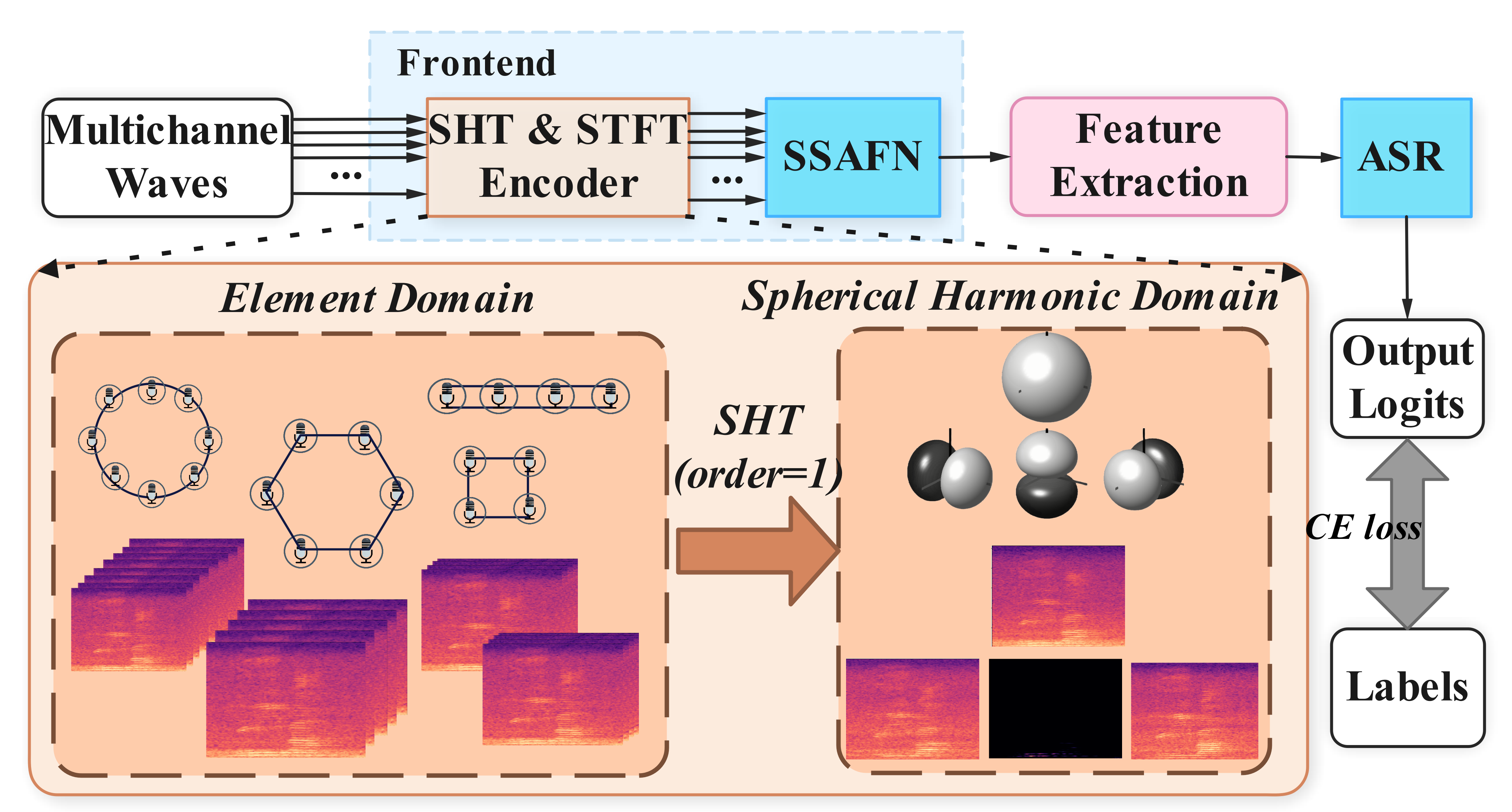}
    \vspace{-0.4cm}
    \caption{SHTNet processing pipeline: Mapping multi-channel signals to the spherical harmonic domain via SHT, followed by spectral enhancement and ASR recognition. The figure illustrates the first-order SHT mapping of element-domain signals from different arrays. Note that during the transformation into the spherical harmonic domain, the planar microphone array geometry restricts vertical sound field sampling, nullifying polar angle-dependent spherical harmonic coefficients. In the figure, this is represented by the spectrograms being black, indicating null values.}
    \label{fig:SHTNet}
    \vspace{-0.5cm}
\end{figure}

\subsection{Architecture Overview}
As illustrated in Fig.~\ref{fig:SHTNet}, the original multi-channel waveforms are transformed into the spherical harmonic domain (SHT) to obtain spatially decoupled spherical harmonic coefficients. The STFT is then applied to each channel to generate time-frequency magnitude spectra. These spectra are enhanced by a Spatial Spectral Attention Fusion Network (SSAFN) to produce a single-channel output, which is converted into Fbank features for ASR recognition. The figure demonstrates the mapping of element-domain signals into the spherical harmonic domain via SHT, showing first-order SHT results. 


\subsection{Frontend}

\subsubsection{SHT \& STFT Encoder}
\vspace{-0.1cm}
The frontend begins by mapping the microphone array signals, which are initially in the element domain, to the spherical harmonic domain via the Spherical Harmonic Transform (SHT). For this purpose, the spherical coordinates of the \(i\)-th microphone are denoted as \((r, \theta_i, \phi_i)\), where \(r\) is the array radius (assuming equidistance of the microphones from the origin), \(\theta_i \in [0, \pi]\) is the polar angle (zenith angle measured from the \(z\)-axis), and \(\phi_i \in [0, 2\pi)\) is the azimuth angle (measured in the \(xy\)-plane from the \(x\)-axis) for the $i$-th microphone.

According to sound field decomposition theory \cite{rafaely2015fundamentals}, the sampled sound pressure \(p(k, r, \theta, \phi)\) in spherical coordinates can be expanded as a series of spherical harmonics:
\begin{align} \label{eq:1}
p(k, r, \theta, \phi) = \sum_{n=0}^{\infty} \sum_{m=-n}^{n} p_{nm}(k, r) Y_n^m(\theta, \phi),
\end{align}
where the wavenumber \( k = 2\pi f/c \), with \( f \) being the frequency and \( c \) the speed of sound in air and the spherical harmonic function \(Y_n^m(\theta, \phi)\) is defined as:
\begin{align}
Y_n^m(\theta, \phi) = \sqrt{\frac{(2n+1)(n-m)!}{4\pi(n+m)!}} P_n^m(\cos\theta) e^{im\phi},
\end{align}
where \(n \in \mathbb{N}\) represents the order, and \(m \in \mathbb{Z}\) satisfies \(-n \leq m \leq n\). \(P_n^m(\cos\theta)\) is the normalized associated Legendre polynomial, which encodes the elevation dependence, while the complex exponential term \(e^{im\phi}\) captures the azimuthal periodicity.
And the spherical harmonic coefficients \(p_{nm}(k, r)\) are given by the following integral \cite{rafaely2015fundamentals}:
\begin{align}
p_{nm}(k, r) = \int_0^{2\pi} \int_0^{\pi} p(k, r, \theta, \phi) \left[ Y_n^m(\theta, \phi) \right]^* \sin\theta \, d\theta \, d\phi.
\end{align}
Under far-field conditions where the distance \(d\) of microphone center and source satisfies \(d > \frac{8r^2f}{c}\) \cite{rafaely2015fundamentals}, the spherical harmonic coefficients can be discretely approximated using an array of \(I\) microphones:
\begin{align}
p_{nm}(k, r) \approx \frac{4\pi}{I} \sum_{i=1}^{I} p(k, \mathbf{r}_i) \left[ Y_n^m(\theta_i, \phi_i) \right]^*,
\end{align}
where \(\mathbf{r}_i = (r, \theta_i, \phi_i)\) represents the spatial coordinates of the \(i\)-th microphone.

The next step involves extracting time-frequency features from the signal using the Short-Time Fourier Transform (STFT). The time-frequency representation, \(P_{nm}(t, f)\), yields the magnitude spectrum, which forms a 3D tensor:
\begin{align}
\boldsymbol{A} \in \mathbb{R}^{C \times T \times F}, \quad A_{nm}(t, f) = \left| P_{nm}(t, f) \right|,
\end{align}
where \(C = (N+1)^2\), represents the number of spherical harmonic channels, \(T\) represents the number of time frames, and \(F\) is the number of frequency bins.
Eq. (\ref{eq:1}) can be approximated for an appropriate finite order $N$ \cite{rafaely2015fundamentals}.

\subsubsection{Spatio-Spectral Attention Fusion Network}
\begin{figure}[t]
  \centering
  \includegraphics[width=0.5\textwidth]{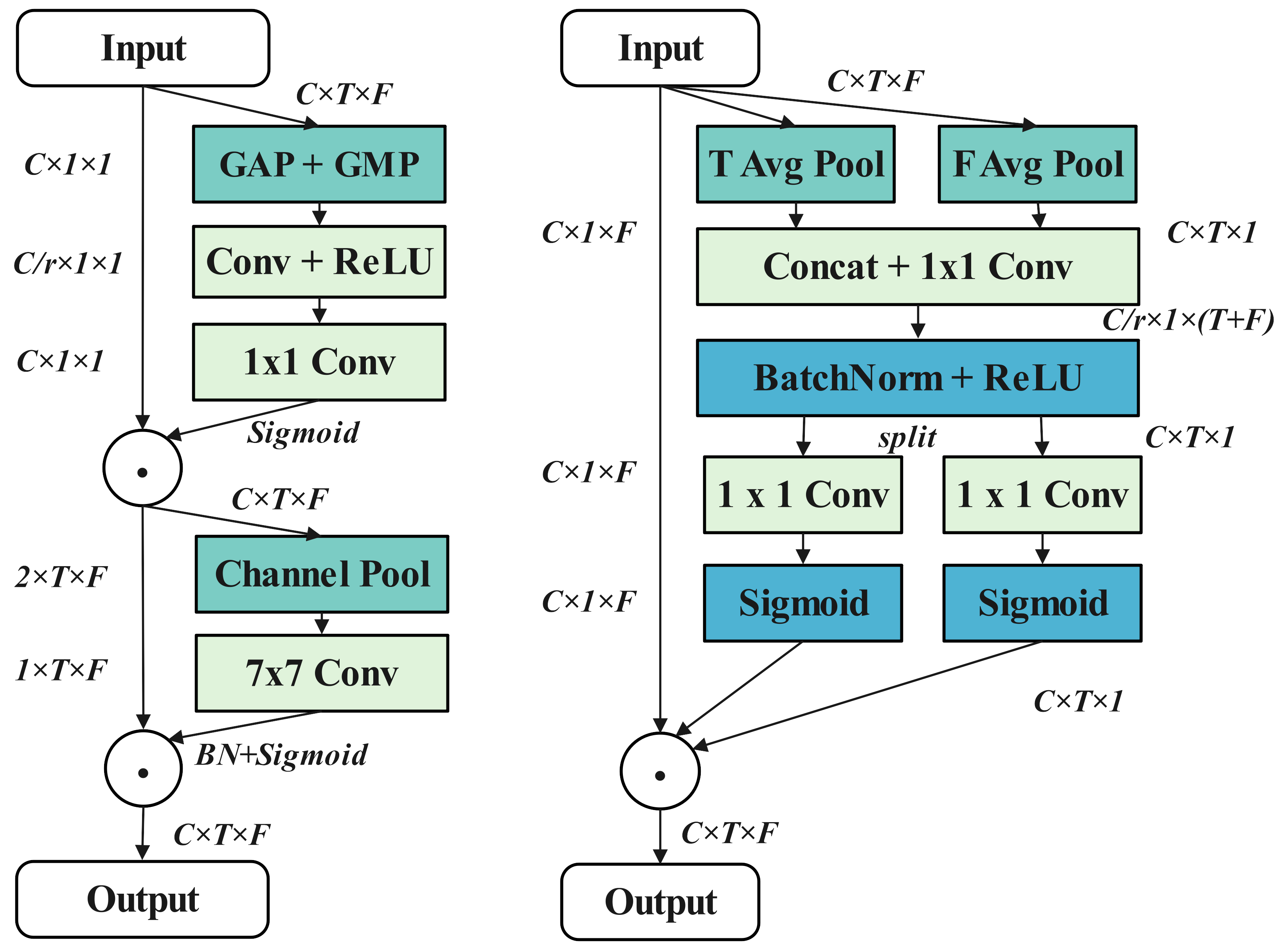}
  \vspace{-0.4cm}
  \caption{Attention modules in SSAFN: CBAM (left) combines spatial and spectral attention, and CoorAttention (right) encodes time/frequency positional relationships. r represents reduction ratio.}
  \label{fig:JointAtt}
  \vspace{-0.5cm}
\end{figure}

The SSAFN employs a hierarchical architecture to progressively refine spatial and spectral features through three key components: spatial and frequency feature extraction, directional filtering, and residual noise suppression.

First, the spatial and frequency feature extraction stage processes the magnitude spectrum  using two cascaded JointAttention blocks.
Each block combines channel and spatial attention mechanisms to enhance feature representation. The block's output \( \mathbf{A}_\text{out} \) is computed as:

\vspace{-0.3cm}
\begin{align}
\begin{split}
\mathbf{A}_\text{out} = \mathbf{A}_\text{in} + \text{CoorAttention}\big(\mathbf{A}_\text{in} + \text{CBAM}(\mathbf{A}_\text{in} + \text{CBAM}(\mathbf{A}_\text{in}))\big).
\end{split}
\end{align}
\vspace{-0.1cm}

The CBAM modules (Left of Fig.~\ref{fig:JointAtt}) apply spatial attention through global average pooling and spectral attention via convolutional layers, while CoorAttention (Right of Fig.~\ref{fig:JointAtt}) enhances spatial dependencies through coordinate-aware positional encoding.
Their sequential application enables progressive refinement from local channel relationships to global coordinate-aware features

Next, the directional filtering stage employs a refined Self-Attention Channel Combinator (rSACC) to amplify target signals. The input \( \mathbf{A}_\text{in} \) first undergoes a log transformation followed by mean-variance normalization (MVN). Then, three linear layers are applied to compute the query \( \mathbf{Q} \), key \( \mathbf{K} \), and value \( \mathbf{V} \) with dimensions of \( T \times C \times E \), \( T \times C \times E \), and \( T \times C \times 1 \), where \( T \) represents the number of time frames, \( C \) is the number of spherical harmonic channels, and \( E \) is the embedding feature dimension.

The attention weights \( \mathbf{w} \) , with dimensions \( T \times C \times 1 \), are calculated as:

\vspace{-0.6cm}
\begin{align}
\mathbf{w} = \text{softmax}\left(\frac{\mathbf{Q}\mathbf{K}^T}{\sqrt{E}}\right) \mathbf{V}
\end{align}
The output of rSACC \( \mathbf{A}_\text{out} \) is obtained by performing a weighted sum of the input \( \mathbf{A}_\text{in} \) with the attention weights \( \mathbf{w} \), summed along the channel axis. The final output tensor \( \mathbf{A}_\text{out} \) has dimensions of \( T \times F \), where \( F \) represents the frequency dimension. This step can be mathematically expressed as:

\vspace{-0.4cm}
\begin{align}
\mathbf{A}_\text{out} = \sum_{C} \mathbf{w} \odot \mathbf{A}_\text{in}
\end{align}

Finally, the residual noise suppression stage applies a multi-head self-attention (MHSA) post-filter to suppress residual noise through multi-spectral correlation analysis. The output is computed as:

\vspace{-0.4cm}
\begin{align}
  \mathbf{A}_\text{out} = \mathbf{A}_\text{in} \cdot \text{MHSA}(\mathbf{A}_\text{in}),
\end{align}

This hierarchical architecture enables SSAFN to effectively integrate spatial and spectral information, achieving robust performance in multi-channel ASR tasks without relying on conventional beamforming techniques.

\subsection{Training Strategy: Joint streaming and non-streaming \& Rand-SHT }
Our training framework is based on the CUSIDE \cite{an22cuside} and CUSIDE-array \cite{kong2024cusidearray} method, which achieves streaming processing by dividing the input into chunks with context. During training, the streaming and non-streaming models share parameters and are jointly trained. The total loss combines cross-entropy objectives from both branches:

\vspace{-0.4cm}
\begin{align}
  \mathcal{L}_{\text{total}} = \mathcal{L}_{\text{non\_stream}} + \mathcal{L}_{\text{stream}},
\end{align}
where \( \mathcal{L}_{\text{non\_stream}} \) and \( \mathcal{L}_{\text{stream}} \) represent the non-streaming and streaming losses, respectively. This dual-objective framework ensures consistency between chunk-wise and full-context outputs while maintaining low-latency performance. 

The Rand-SHT training strategy randomly selects \( I'\) (\( 2 \leq I' \leq I \)) microphones during training and recalculates spherical harmonic coefficients:

\vspace{-0.4cm}
\begin{align}
    \tilde{p}_{nm}(k, {r}) \approx \frac{4\pi}{I'} \sum_{i \in I'} p(k, \mathbf{r}_i') \left[Y_n^m(\theta_i', \phi_i')\right]^*,
\end{align}
where \( (\theta_i', \phi_i') \) are computed based on the relative positions of the selected microphones. This approach simulates diverse array geometries during training, enhancing robustness to array variability.

\vspace{-0.4cm}
\section{Experiments}
\vspace{-0.1cm}

\subsection{Datasets and Evaluation Metrics}
\vspace{-0.1cm}

Datasets include:
\vspace{-0.1cm}
\begin{itemize}
    \item \textit{AISHELL-4} \cite{fu2021aishell4}: \textbf{ID test} with 43.4h training / 2.3h validation / 8.9h test of non-overlapping meeting speech with 8-channel arrays. Testing uses channels 1,3,5,7 (4ch square array) and 1,5 (2ch binary array).
    \item \textit{Alimeeting} \cite{yu2022alimeeting}: \textbf{OOD test} of conference-style non-overlapping segments (3.6h test / 1.2h eval) from M2MeT challenge.
    \item \textit{XMOS} \cite{kong2024cusidearray}: \textbf{OOD test} with 10ch real meeting recordings in noisy environments ($\sim$40 utterances), collected using an XMOS microphone array board.
\end{itemize}
\vspace{-0.1cm}
Evaluations focus on character error rate (CER) under ID and OOD scenarios, and computational efficiency (front-end GFLOPS and streaming decoding time on NVIDIA 3090 GPU).

\vspace{-0.1cm}
\subsection{Experiment Setup}
\vspace{-0.1cm}
\subsubsection{Training Details}
\vspace{-0.1cm}
Our models are trained with the CAT ASR toolkit \cite{An2020CATAC}.
Different models are trained on AISHELL-4's train set, measured by character error rate (CERs).
The SHT order \( N \) is set to 4 , as preliminary experiments have demonstrated that this value optimally balances spatial resolution and computational efficiency.
Training uses a 16 kHz sampling rate with 512-point STFT parameters (25 ms frame length, 10 ms shift) to extract 80-dimensional log-Mel filterbank features.  Optimization leverages the Adam optimizer with Transformer learning rate scheduler, gradient clipping, and early stopping when validation loss drops below \( 10^{-6} \).

Streaming signal processing uses 400ms chunks with 800ms left context and 400ms right context (used 50\% randomly during training, omitted during testing). Chunk size is dynamically sampled from 350-450ms in training.

\vspace{-0.1cm}
\subsubsection{Network Details}
\vspace{-0.1cm}
The ASR encoder adopts a 12-layer Conformer architecture with a CTC objective, configured with 4 attention heads, 256-dimensional attention layers, and 3038-dimensional feedforward layers.

For SHTNet, the frontend integrates four CBAM modules with spatial attention kernels (sizes 9, 7, 5, 3) for multi-scale spatial patterns, and spatial attention reduction is set to 5. The CoorAttention module follows the same reduction strategy. All modules maintain consistent input/output channel dimensions (25 channels) except rSACC, and the MHSA component uses 2 attention heads with 64-dimensional attention layers.

In comparative experiments, the CUSIDE-Array baseline uses a three-layer BLSTM with 320 hidden units per direction for complex mask estimation. 

\vspace{-0.2cm}

\begin{figure}[ht]
  \centering
  \vspace{-0.3cm}
  \includegraphics[width=1\linewidth]{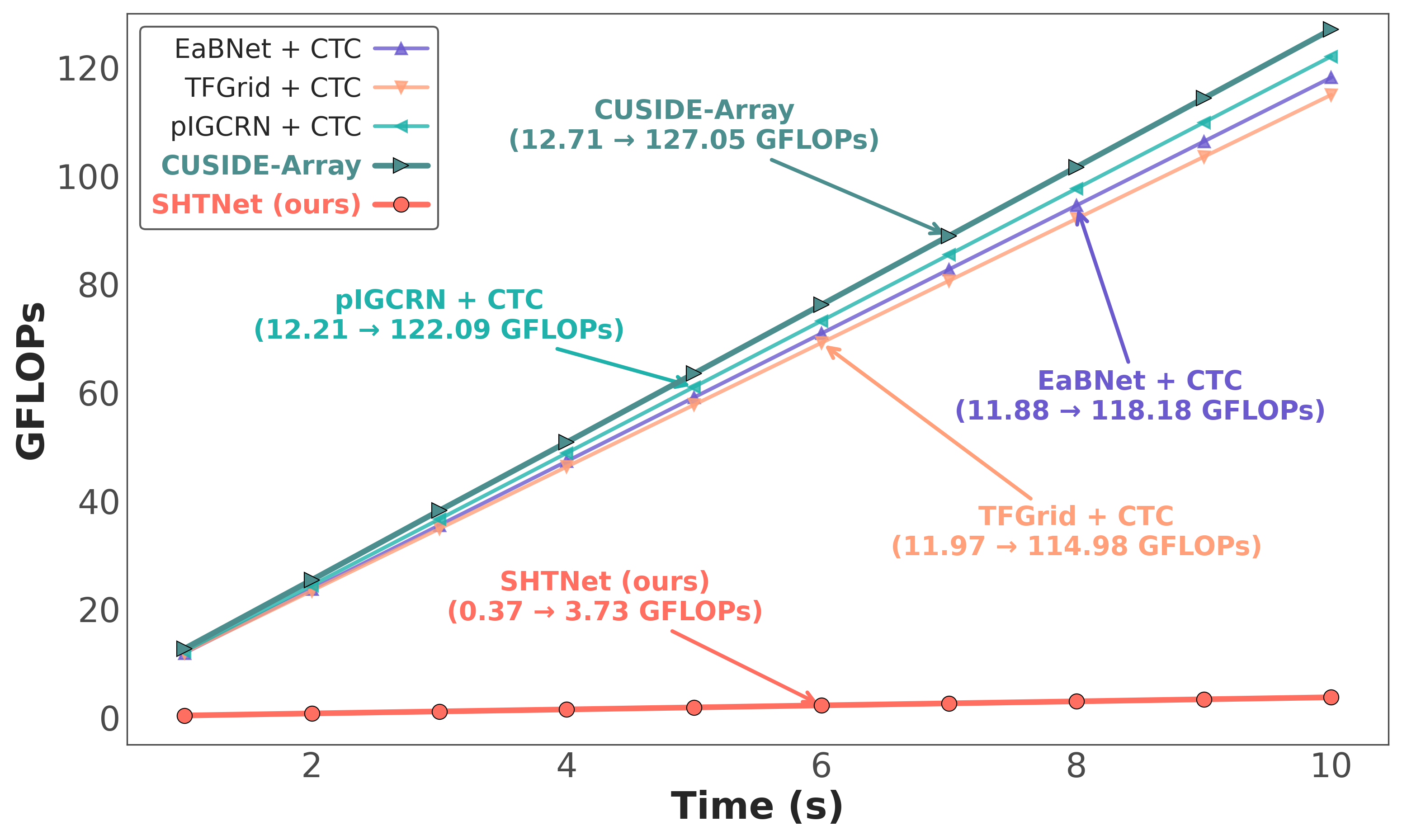}
  \caption{Comparison of front-end GFLOPs Among Different Models with Input Lengths Ranging from 1s to 10s}
  \label{fig:gfps}
  \vspace{-0.3cm}
\end{figure}

Other baselines (EaBNet, TFGridNet, pIGCRN) are fine-tuned end-to-end with the pre-trained ASR encoder from AISHELL-4. To uniform GPU memory allocation during training for fair comparison, TFGridNet employs a single-layer LSTM (128 hidden units), 256-dimensional query/key projections, and 16-dimensional embeddings, while pIGCRN reduces embedding channels to 20 while preserving its architecture.

\vspace{-0.3cm}
\section{Results and Discussions}


\vspace{-0.3cm}
\subsection{Non-Streaming Experiments}
\vspace{-0.2cm}

Table~\ref{tab:nonstreaming} and Figure~\ref{fig:gfps} collectively reveal two critical observations about our framework. \textbf{First}, SHTNet achieves the lowest average CER (39.26\%) across all test configurations while maintaining \textbf{ultra-low computational costs} (3.73 GFLOPS for 10s inputs). This lightweight superiority manifests in two aspects: 1) Compared to conventional beamforming approaches, it reduces frontend parameters by 92.4\% (0.38M vs. CUSIDE-Array's 5.00M); 2) Against neural baselines like TFGridNet+CTC (0.39M/42.93\% CER) and pIGCRN+CTC (0.81M/44.58\% CER), it achieves superior accuracy with comparable or fewer parameters. The 97.1\% GFLOPS reduction (Figure~\ref{fig:gfps}) stems from SHT's geometry-agnostic spatial encoding, which eliminates redundant multi-channel computations via spherical harmonic decomposition. This innovation enables the modeling of spatial-spectral relationships using a simple attention network, thus avoiding the computational overhead associated with LSTM architectures, which are used in CUSIDE-Array.

\textbf{Second}, SHTNet demonstrates exceptional \textbf{cross-configuration robustness}. When microphone channels decrease from 8 to 2, CER increases by merely 2.32\% (29.34\%→31.66\%), significantly outperforming CUSIDE-Array's 6.55\% degradation (29.65\%→36.20\%). This stability extends to cross-domain scenarios: on real-world XMOS dataset, SHTNet achieves 74.85\% CER – 4.06\% lower than CUSIDE-Array and 22.18\% better than pIGCRN+CTC, with similar performance gains observed on AliMeeting dataset. Ablation studies reveal the source of this robustness: 1) Removing Rand-SHT training causes XMOS CER to surge by 22.08\% (74.85\%→96.93\%) and degrades performance across various array configurations, particularly under 2-channel setups (31.66\%→38.58\%); 2) The impact of disabling other modules proves  less pronounced compared to Rand-SHT removal. These findings confirm that Rand-SHT training effectively enhances the SHT-based model's robustness to diverse array geometries. The improvement of SHTNet over the suboptimal CUSIDE-Array in nonstreaming ASR is significant, with a p-value of 5.71e-8 by matched-pair significance test \cite{significance}.

\begin{table}[h!]
    \centering
    \caption{CER Results of Non-Streaming Models. The ASR architecture is identical across all models, with a size of 20.77M. All single-channel models use channel 0 as the input. EaBNet, TFGridNet, and pIGCRN perform end-to-end fine-tuning on the single-CTC, which serves as the pre-trained ASR encoder from AISHELL-4. }
    \vspace{-0.3cm}
    \label{tab:nonstreaming}
    \setlength{\tabcolsep}{2pt} 
    \renewcommand{\arraystretch}{1.1} 
    \small 
    \begin{adjustbox}{max width=\columnwidth}
    \begin{tabular}{l|c|ccc|cc|c|c}
    \toprule
    \multirow{2}{*}{Model} & \multirow{2}{*}{\makecell{Para. of\\FE (M)}} & \multicolumn{3}{c|}{Aishell-4} & \multicolumn{2}{c|}{Alimeeting} & \multicolumn{1}{c|}{XMOS} & \multirow{2}{*}{Avg.} \\
    & & 8-ch & 4-ch & 2-ch & test & eval & test & \\
    \midrule
    CUSIDE \cite{an22cuside}          & --    & 35.22 & 35.22 & 35.22 & 40.78 & 45.42 & 84.36 & 46.20 \\
    CUSIDE-Array \cite{kong2024cusidearray}    & 5.00  & 29.65 & 32.01 & 36.20 & 34.62 & 38.47 & 78.91 & 41.64 \\
    single-CTC      & --    & 35.65 & 35.65 & 35.65 & 40.27 & 44.10 & 87.13 & 46.41 \\
    EaBNet \cite{li2022eabnet} + CTC    & 3.21  & 31.79 & 35.11 & 36.97 & 35.46 & 39.32 & 93.56 & 45.37 \\
    TFGrid \cite{wang2023tf} + CTC    & 0.39  & 30.51 & 33.44 & 36.61 & 33.88 & 37.18 & 85.94 & 42.93 \\
    pIGCRN \cite{pan2024efficient} + CTC & 0.81  & 31.20 & 32.32 & 33.42 & 35.03 & 38.47 & 97.03 & 44.58 \\
    \midrule
    \textbf{SHTNet(ours)}  & 0.38  & \textbf{29.34} & \textbf{29.57} & \textbf{31.66} & \textbf{33.14} & \textbf{37.01} & \textbf{74.85} & \textbf{39.26} \\
    \hspace{2mm}w/o Rand-SHT &  0.38 & 30.34 & 32.35 & 38.58 & 34.46 & 37.76 & 96.93 & 45.07 \\
    \hspace{2mm}w/o SHT     & 0.38  & 32.18 & 32.29 & 33.49 & 35.69 & 38.53 & 98.12 & 45.05 \\
    \hspace{2mm}w/o JointAtt & 0.37 & 30.49 & 30.87 & 32.60 & 33.76 & 37.48 & 97.23 & 43.74 \\
    \hspace{2mm}w/o rSACC & 0.31 & 32.05 & 32.26 & 34.09 & 34.77 & 38.92 & 92.18 & 44.05 \\
    \hspace{2mm}w/o MHSA & \textbf{0.06} & 30.20 & 30.38 & 32.21 & 33.89 & 37.78 & 98.12 & 43.76 \\
    \bottomrule
    \end{tabular}
    \end{adjustbox}
    \vspace{-0.3cm}
\end{table}

\vspace{-0.1cm}

\begin{table}[ht]
  \centering
  \caption{Streaming CER (\%) and Latency Results. We tested the decoding time of different models on the AISHELL-4 8-ch test set with the same chunk size (400ms) on a single 3090 GPU, recorded as Time(s). {$\Delta t$(s)} represents the total decoding latency introduced by the addition of the frontend.}
  \vspace{-0.4cm}
  \label{tab:streaming_results}
  \setlength{\tabcolsep}{2pt} 
    \renewcommand{\arraystretch}{1.1} 
    \small 
    \begin{adjustbox}{max width=\columnwidth}
    \begin{tabular}{l|cc|ccc|cc|c|c}
    \toprule
     \multicolumn{1}{c|}{\multirow{2}{*}{Model}} & \multirow{2}{*}{\makecell{Time(s)}} & \multirow{2}{*}{\makecell{$\Delta t$(s)}} & \multicolumn{3}{c|}{Aishell-4} & \multicolumn{2}{c|}{Alimeeting} & \multicolumn{1}{c|}{XMOS} & \multirow{2}{*}{Avg} \\
    & & & 8-ch & 4-ch & 2-ch & test & eval & test & \\
    \midrule
    CUSIDE         & \textbf{722.83} & -- & 41.01 & 41.01 & 41.01 & 46.82 & 51.40 & 86.63 & 51.31 \\
    CUSIDE-Array   & 978.26 & 255.43 & 35.08 & 37.11 & 41.54 & 39.53 & 43.08 & 82.08 & 46.40 \\
    \textbf{SHTNet(ours)}   & 820.25 & \textbf{97.42} & \textbf{34.37} & \textbf{34.74} & \textbf{36.80} & \textbf{37.66} & \textbf{41.61} & \textbf{79.50} & \textbf{44.11} \\
    \bottomrule
  \end{tabular}
  \end{adjustbox}
  \vspace{-0.4cm}
\end{table}

\vspace{-0.3cm}
\subsection{Streaming Experiments}
\vspace{-0.1cm}
As shown in Table~\ref{tab:streaming_results}, our streaming implementation offers two key advantages.
First, it achieves real-time efficiency without compromising accuracy. When processing 8-channel inputs from the Aishell-4 test set, SHTNet significantly reduces frontend processing latency to 97.42s\footnote{This refers to total processing time for the 8.9h test set. Per-utterance latency is 15.5ms.} compared to CUSIDE-Array's 255.43s.
This represents a 62\% reduction in additional computational overhead under identical device and chunk-size configurations, while simultaneously achieving superior recognition accuracy (34.37\% CER vs 35.08\%). Second, the framework maintains robust performance under streaming constraints. When handling dynamically varying channel counts, SHTNet exhibits only a 2.4\% CER degradation (34.37\%→36.80\%), substantially outperforming CUSIDE-Array's 6.46\% performance drop (35.08\%→41.54\%).

This advantage is driven by SHT's enhanced spatial information extraction and attention-based architecture in SHTNet, which reduces sensitivity to time-series duration, unlike traditional LSTM models. Together, they enable robust performance in real-world streaming processing.

\vspace{-0.2cm}
\section{Conclusion and Future Work}
\vspace{-0.1cm}
We presents a lightweight and robust multi-channel speech recognition framework, SHTNet, leveraging SHT for efficient spatial modeling. SHTNet demonstrates excellent performance in both ID and OOD tasks, achieving low error rates and computational efficiency. Future work will focus on optimizing the system for real-time deployment, exploring self-supervised learning for better cross-domain adaptation, and extending the model to multi-speaker scenarios.

\bibliographystyle{IEEEtran}
\bibliography{mybib}

\end{document}